\documentclass[10pt]{iopart}
\usepackage{iopams}
\usepackage{epsfig,cite}
\newcommand{\be}{\begin{equation}}
\newcommand{\ee}{\end{equation}}
\newcommand{\bd}{\begin{displaymath}}
\newcommand{\ed}{\end{displaymath}}
\newcommand{\BE}{\begin{eqnarray}}
\newcommand{\EE}{\end{eqnarray}}

\newcommand{\sgn}{{\rm sgn}}

\newcommand{\id}{{\rm 1\!\! I}}

\newcommand{\bu}{\ensuremath{\mathbf{u}}}

\newcommand{\bR}{\ensuremath{\mathbf{R}}}

\newcommand{\avg}[1]{\left\langle{#1}\right\rangle}

\begin{document}

\title{Spherical grand-canonical minority games with and without score discounting}

\author{Alex Bladon, Thomas Richardson and Tobias Galla}

\address{The University of Manchester, School of Physics and Astronomy, Manchester M13 9PL, United Kingdom}

\begin{abstract}
We present a spherical version of the grand-canonical minority game
(GCMG), and solve its dynamics in the stationary state. The model displays several types of transitions between multiple ergodic phases and one non-ergodic phase. We derive analytical solutions, including exact expressions for the volatility, throughout all ergodic phases, and compute the phase behaviour of the system. In contrast to conventional GCMGs, where the introduction of memory-loss precludes analytical approaches, the spherical model can be solved also when exponential discounting is taken into account. For the case of homogeneous incentives to trade $\varepsilon$ and memory loss rates $\rho$, an efficient phase is found only if $\rho=\varepsilon=0$. Allowing for heterogeneous memory-loss rates we find that efficiency can be achieved as long as there is any {\em finite} fraction of agents which is not subject to memory loss.

\end{abstract}


\ead{\tt alex.bladon@postgrad.manchester.ac.uk, thom.j.richardson@googlemail.com, tobias.galla@manchester.ac.uk}

\section{Introduction}
Let's face it: more than ten years after its introduction, the
theoretical understanding of the minority game (MG)
\cite{cz,book1,book2} is still only partial, and several questions
regarding its statistical mechanics remain open and pose interesting
challenges for theoretical physicists. While analytical solutions of
the conventional MG, valid in its ergodic phase, have now been
obtained based on static and dynamical techniques from spin glass
physics \cite{book1,book2,hc,rev}, the analysis of the non-ergodic
phase is still far from being complete. Furthermore, recent numerical
observations in grand-canonical MGs (GCMG) with finite score-memories
\cite{finitemem} have indicated anomalous fluctuations, similar to
stylised facts observed in time series of real-world financial
markets. Such stylised facts are observed in GCMGs without memory-loss
as well \cite{book1,book2,gcmg}, but score discounting introduces
other hitherto unknown dynamical effects \cite{finitemem}. Such models
with finite score memory resist analytical approaches, as solutions
rely on the existence of so-called `frozen' agents, i.e. agents who
never switch strategies over the course of the learning dynamics of
the MG. No such agents can be found in the case of finite
score-memories, and the understanding of MGs with exponential
discounting, along with obtaining solutions in of MGs in their
non-ergodic phases, has hence been pointed out as one of the remaining
challenges in MG theory \cite{finitemem}.

The purpose of our work is to present and to solve a spherical version
of the GCMG without and with finite score-memories. Our approach here follows a
strategy well known in statistical mechanics, namely to replace
Ising-like discrete degrees of freedom of a spin model by continuous
ones, subject to an overall spherical constraint. While the static
partition function of an Ising-like system is a sum over corners of a
high-dimensional hypercube, this sum is effectively replaced with an
integral over the surface of a sphere (passing through the corners of
the hypercube) in spherical spin models. This procedure dates back to
1952, when Berlin and Kac devised a spherical version of the Ising
model \cite{BerlinKac}, solvable in any dimension, and has been
carried out in the context of various spin models since then, see for
example \cite{Crisanti} for spherical versions of models of spin
glasses, \cite{sphericalhopfield} for a spherical version of the
Hopfield model of neural networks and \cite{GallCoolSher03,GallSher05,PapaCool}
for earlier studies of spherical minority games. The construction of spherical MGs admittedly reduces the complexity of the model considerably, and, due to the effectively harmonic update rules, only two modes contribute to macroscopic dynamics in the stationary state, fully frozen and fully oscillatory terms. However, despite this simplicity, spherical MGs have been seen to generate phase behaviour similar to that of conventional MGs, in particular a phase transition between ergodic and non-ergodic phases of the same type as in non-spherical MGs is found. The simplicity of the dynamical trajectories in spherical MGs might preclude the observation of stylised facts and non-Gaussian fluctuations, but this reduction in complexity is compensated for by the full analytical tractability allowing for an exact computation of the market volatility, which is as yet not possible in conventional MGs. Spherical MGs might therefore serve as a fully solvable starting point to which non-linear effects might be added in a perturbative fashion, maintaining analytical tractability as far as possible \cite{marsili}. Spherical MGs with memory-loss have briefly been considered in \cite{DeMaGall}, but to our knowledge no systematic study of the detailed phase behaviour and ergodicity properties of spherical grand-canonical MGs has been carried out.

The paper is based on \cite{alexthom} and structured as follows: we
will first introduce the spherical GCMG in Sec. \ref{sec:def} and then
derive a self-consistent macroscopic theory in the following
section. In particular we will obtain analytical expressions for the
order parameters of the system in its ergodic phases. In
Sec. \ref{sec:pg} we then discuss the phase behaviour of the model
without memory loss in detail. Spherical GCMG models with homogeneous
and with heterogeneous memory-loss rates are studied in
Sec. \ref{sec:finitemem}. Throughout Sections \ref{sec:pg} and
\ref{sec:finitemem} we successfully test our theoretical findings
against simulations. A summary and outlook is given in the concluding
section.

\section{Definition of the model}\label{sec:def}
The GCMG describes a population of $N=N_s+N_p$ agents, $N_p$ of which are so-called `producers', and the remaining $N_s$ are usually referred to as `speculators'. For simplicity
we will assume that the speculators carry the labels $i=1,\dots,N_s$, and that the producers are labelled by $i=N_s+1,\dots,N$. All agents are presented with a piece of
information $\mu(t)\in\{1,\dots,P\}$ randomly chosen at each step of the game. $\mu(t)$ here encodes information on the `state of the world', on which the agents react. The
mathematically relevant regime is the case in which the number of possible information patterns is proportional to the number of players in the game, i.e. $P=\alpha N$, with $\alpha\sim{\cal O}(N^0)$. Following the common conventions we will abbreviate $n_s\equiv N_S/P$ and $n_p\equiv N_P/P$, resulting in $\alpha=(n_s+n_p)^{-1}$. Each agent in the GCMG is then equipped with one trading strategy, providing a map from the space of all possible information patterns onto the set of binary actions $\{-1,+1\}$. We will write $\bR_i\equiv(R_i^1,\dots,R_i^{\alpha N})\in\{-1,1\}^{\alpha N}$ in the following for agent $i$'s trading strategy. Being presented with information pattern $\mu$ agent $i$'s trading action, when using his strategy, is hence $R_i^\mu$. While producers in the GCMG trade no matter what (i.e. they use their trading strategy at each time-step), speculators are given the option to use their strategy vector or to abstain. Whether they trade or not depends on their valuation of their active strategy, usually based on a score variable $u_i(t)$ they assign to their strategy. Speculator $i$ then trades at time $t$ if and only if $u_i(t)>0$. The total action at time $t$ can hence be written as
\be
A^{\mu(t)}[\bu(t)]=N^{-1/2}\left(\sum_{i=1}^{N_s}R_i^{\mu(t)}n_i(t)+\sum_{i=N_s+1}^{N}R_i^{\mu(t)}\right),
\ee
where $n_i(t)=\Theta[u_i(t)]$, with $\Theta(\cdot)$ the step function, i.e. $\Theta(x)=1$ if $x>0$ and $\Theta(x)=0$ otherwise. In the batch GCMG score updates of speculators are then performed according to the following learning rule\footnote{We here note that different ways of introducing memory-loss into the learning dynamics have been proposed in the literature. In particular \cite{finitemem} consider a dynamics in which the term proportional to the $J_{ij}$ carries a pre-factor $\rho$. We here restrict the analysis to the process (\ref{eq:update}), which is consistent with \cite{finitemem2} and has the advantage of reproducing the standard GCMG in the limit $\rho=0$. See also \cite{mosetti} for further comparison of the two types of dynamics in some special cases.} \cite{finitemem2,finitemem}, see also \cite{book1,book2} for further details of the batch MG:
\be\label{eq:update}
u_i(t+1)=(1-\rho)u_i(t)-\frac{2}{\sqrt{N}}\sum_{\mu} R_i^\mu A^{\mu}[\bu(t)]-\varepsilon,
\ee
where $0\leq \rho\leq 1$ accounts for effects of memory-loss or equivalently corresponds to exponential discounting over time \cite{finitemem,finitemem2}. The model parameter $\varepsilon$ is a disincentive to trade, or equivalently a marginal payoff agents need to make in order to consider the option of trading \cite{gcmg}. Eq. (\ref{eq:update}) can be written as
\be\label{eq:update2}
u_i(t+1)=(1-\rho) u_i(t)-\sum_j J_{ij} n_j(t)-\varepsilon,
\ee
with
\be
J_{ij}=\frac{2}{N}\sum_\mu R_i^\mu R_j^\mu.
\ee
Writing $n_i(t)=\frac{1}{2}(1+\sgn[u_i(t)])$ a spherical version of these dynamics can be introduced by the replacement $n_i(t)\to\frac{1}{2}(1+\phi_i(t))$ where $\phi_i$ is now a continuous degree of freedom and where the vector $\boldmath{\phi}(t)=(\phi_1(t),\dots,\phi_N(t))$ fulfills the spherical constraint
\be
\frac{1}{N}\sum_i \phi_i^2(t)=1,
\ee
at all times. Following \cite{GallSher05} the $\{\phi_i(t)\}$ are given by
\be
\phi_i(t)=\frac{u_i(t)}{\lambda(t)},
\ee
with a suitable $\lambda(t)$, so that the spherical constraint is respected. 

The update rule of the spherical GCMG then reads
\be
u_i(t+1)=(1-\rho) u_i(t)-\frac{1}{2}\sum_j J_{ij} (1+\phi_j(t))-\varepsilon,
\ee
and will provide the starting point for the further analysis. Potential ergodicity breaking (i.e. sensitivity to initial conditions) will be probed by starting the dynamics from initial conditions with differently biased strategy scores. In particular it is common to distinguish between so-called `tabula rasa' starts at low initial scores and highly biased starts. The spherical constraint excludes a starting point from zero-initial conditions ($u_i(t=0)=0~\forall i$). In our simulations we will use starts of the form $u_i(0)=u_0$, with $u_0>0$. Small values of $u_0$ (e.g. $u_0=0.01$) then corresponds to mostly unbiased starts, whereas a score bias can be imposed by choosing larger values of $u_0$. 

We will in particular be concerned with behaviour of the so-called predictability and volatility of the model market defined by the MG. These are commonly studied quantities in MG, we will here only give a brief definition of these observables, further details regarding the economic background can be found in \cite{book1,book2}. The so-called predictability $H$ measures the degree to which the overall bid $A^\mu(t)$ can statistically be predicted given the occurrence of a given information pattern $\mu$. Specifically one defines
\be
H=\frac{1}{P}\sum_{\mu=1}^P\left[\lim_{T\to\infty}T^{-1}\sum_{t\leq T} A^\mu(t)\right]^2.
\ee
Each individual square bracket is non-zero if and only if the overall bid is predictable given $\mu$, hence $H>0$ indicates that the time-series of bids generated by the MG dynamics contains elements of predictability. $H=0$ on the other hand corresponds to an unpredictable or so-called symmetric phase. The second key observable is the volatility, defined as
\be
\sigma^2=\lim_{T\to\infty}\frac{1}{PT}\sum_{t\leq T}\sum_\mu A^\mu(t)^2,
\ee
where averages over the disorder and potentially random initial conditions of the dynamics are implied in these expressions, see e.g. \cite{book2} for details. One further observable we will be interested in is the `traded volume'
\be
V(t)=\lim_{T\to\infty}T^{-1}\sum_{t\leq T}\frac{1}{N}\sum_i\frac{1+\phi_i(t)}{2}.
\ee
We here note that $\phi_i(t)=-1$ corresponds to agent $i$ abstaining at time-step $t$ ($n_i(t)=0$), whereas $\phi_i(t)=1$ represents a situation in which $n_i(t)=1$, i.e. agent $i$ trades one unit at time $t$. We also point out that $0\leq V(t)\leq 1$, which can be seen from $\left|N^{-1}\sum_i \phi_i(t)\right|\leq 1$. The latter in turn is due to the inequality $\left|N^{-1}\sum_i \phi_i(t)\right|^2\leq N^{-1}\sum_i \phi_i(t)^2$ and the spherical constraint. The statistical mechanics analysis will eventually be concerned with the computation of these observables in the thermodynamic limit, $N\to\infty$, at finite control parameters $\alpha,n_s$ and $n_p$.
\section{Generating functional analysis}
\subsection{Macroscopic theory}
The analysis of MG is conveniently carried out based on a path-integral approach\cite{dedom}, leading to an effective self-consistent theory for the relevant macroscopic order parameters. We will here not report the details of the mathematical steps, as these have been discussed for similar models at length in the literature \cite{hc,book2}. The outcome of this procedure applied to the spherical GCMG is the following stochastic process for an effective speculator
\be\label{eq:effproc}
u(t+1)=(1-\rho)u(t)-\varepsilon-\alpha\sum_{t'}(\id+G)^{-1}_{tt'}(1+\phi(t'))+\sqrt{\alpha}\eta(t),
\ee
where $\eta(t)$ is Gaussian noise of zero average, and where we have
\BE
C_{tt'}&=&f_s\avg{\phi(t)\phi(t')}+f_p,\\
G_{tt'}&=&f_s\frac{\delta}{\sqrt{\alpha}\delta \eta(t')}\avg{\phi(t)},\\
M_t&=&f_s\avg{\phi(t)}+f_p,\\
\Lambda_{tt'}&\equiv&\avg{\eta(t)\eta(t')}=[(\id+G)^{-1}D(\id+G^T)^{-1}]_{tt'},\label{eq:scnoise}\\
D_{tt'}&=&1+C_{tt'}+M_t+M_{t'},
\EE
as well as $\phi(t)=u(t)/\lambda(t)$.
The notation $\avg{\cdots}$ here refers to an average over realisations of the stochastic process (\ref{eq:effproc}), and we have introduced the fraction of speculators, $f_s=n_s/(n_s+n_p)$, and the fraction of producers, $f_p=1-f_s$, for later convenience. Producers are always frozen, i.e. we have $\phi_i(t)=1$ for $i=N_s+1,\dots,N$ at all times, hence they contribute a constant $f_p=n_p/(n_s+n_p)$ to the total magnetization $M=\lim_{N\to\infty}N^{-1}\sum_{i=1}^N \overline{\phi_i(t)}$ and to the correlation function $C(t,t')=\lim_{N\to\infty} N^{-1}\sum_{i=1}^N \overline{\phi_i(t)\phi_i(t')}$, and have no effect on the response function $G(t,t')$ ($\overline{\cdots}$ here denotes an average over the disorder, i.e. over the random assignments of strategy tables at the beginning of the game). We will explicitly separate the contributions from the group of speculators from those of the producers in the order parameters below, and will write
\BE
M(t)&=&f_sM_s(t)+f_p\label{eq:mm},\\
C(t,t')&=&f_sC_s(t,t')+f_p\label{eq:cc},\\
G(t,t')&=&f_sG_s(t,t')\label{eq:gg},
\EE
where quantities with subscript $s$ are from the group of speculators.

Replacing $u(t)$ with $\lambda(t)\phi(t)$ in (\ref{eq:effproc}) and similarly for $u(t+1)$ we can write the effective process as
\BE
\lambda(t+1)\phi(t+1)&=&\lambda(t)(1-\rho)\phi(t)-\varepsilon-\alpha\sum_{t'}(\id+G)^{-1}_{tt'}\nonumber\\
&&-\alpha\sum_{t'}(\id+G)^{-1}_{tt'}\phi(t')+\sqrt{\alpha}\eta(t).\label{eq:effpr}
\EE
Following the steps outlined in \cite{GallSher05} the following three equations can then be derived for the time-evolution of the dynamical order parameters $C_s(t,t'), G_s(t,t'), M_s(t)$:
\BE
\lambda(t+1)M_s(t+1)&=&\lambda(t)(1-\rho)M_s(t)-\varepsilon-\alpha\sum_{t'}(\id+G)^{-1}_{tt'}\nonumber \\
&&-\alpha\sum_{t'}(\id+G)^{-1}_{tt'}M_s(t'),\label{eq:1}\\
\lambda(t+1)C_s(t+1,t')&=&\lambda(t)(1-\rho)C_s(t,t')-\varepsilon M_s(t')\nonumber\\
&&-\alpha M_s(t')\sum_{t''}(\id+G)^{-1}_{tt''}\nonumber\\
&&
-\alpha\sum_{t''}(\id+G)^{-1}_{tt''}C_s(t',t'')+\alpha\left[\Lambda G_s^T\right]_{tt'},\label{eq:2}\\
\lambda(t+1)G_s(t+1,t')&=&\lambda(t)(1-\rho)G_s(t,t')\nonumber \\
&&-\alpha\sum_{t''}(\id+G)^{-1}_{tt''}G_s(t'',t')+\delta_{tt'}\label{eq:3}.
\EE
Eqs. (\ref{eq:mm},\ref{eq:cc},\ref{eq:gg}) allow for an equivalent formulation in terms of $C,G$ and $M$. The analysis of these equations proceeds along the lines of \cite{GallCoolSher03,GallSher05}. Ergodic stationary phases can here be addressed making the usual assumption of time-translation invariance, $C_s(t,t')=C_s(t-t'), G_s(t,t')=G_s(t-t'), M_s(t)=M_s$, and of finite integrated response, $\chi_s\equiv\sum_{\tau} G_s(\tau)<\infty$, where $\tau=t-t'$ . We note here that $\chi_s<\infty$ is equivalent to  $\chi\equiv\sum_{\tau} G(\tau)<\infty$. Guided by computer simulations one identifies three types of ergodic phases, one with unbounded frozen dynamics and two with bounded motion. We will address these different phases separately in the following, and derive explicit closed-form results for the relevant order parameters in the different regimes.

\subsection{Frozen ergodic phase with unbounded dynamics (UF)}
A phase with unbounded trajectories of score valuations $u_i(t)$ is found in simulations at vanishing memory loss $\rho=0$. In this phase one finds $\lambda(t)=\lambda_1 t$ at stationarity in numerical simulations, and observes that all speculators follow trajectories of the type $u_i(t)=v_it$, at constant $\phi_i(t)\equiv \phi_i=v_i/\lambda_1$. One thus has $C(t,t')\equiv 1$ for all $t,t'$. Making a corresponding ansatz on the level of the effective process, after averaging over time one has
\be
v=-\varepsilon-\frac{\alpha}{1+\chi}-\frac{\alpha v}{\lambda_1(1+\chi)}+\sqrt{\alpha}\eta,
\ee
i.e
\be
v(\eta)=\left[1+\frac{\alpha}{\lambda_1(1+\chi)}\right]^{-1}\left(-\varepsilon-\frac{\alpha}{1+\chi}+\sqrt{\alpha}\eta\right).
\ee
$\eta$ in this ansatz is a {\em static} Gaussian random variable of zero mean and with variance prescribed by the self-consistency relation (\ref{eq:scnoise}):
\be
\avg{\eta^2}=\frac{2(1+M)}{(1+\chi)^2}.
\ee
Self-consistency furthermore requires
\be
\lambda_1=\sqrt{\avg{v^2}}, ~~ M=f_s\frac{1}{\lambda_1}\avg{v}+f_p, ~~ \chi=\frac{f_s}{\lambda_1\sqrt{\alpha}} \avg{\frac{\partial v}{\partial\eta}},
\ee
and these averages are readily carried out given the explicit form of $v(\eta)$ and taking into account the statistics of $\eta$. This gives
\BE
\lambda_1&=&\left[1+\frac{\alpha}{\lambda_1(1+\chi)}\right]^{-1}\left[\left(\varepsilon+\frac{\alpha}{1+\chi}\right)^2+\alpha\frac{2(1+M)}{(1+\chi)^2}\right]^{1/2},\label{eq:uf1} \\
M&=&\frac{f_s}{\lambda_1}\left[1+\frac{\alpha}{\lambda_1(1+\chi)}\right]^{-1}\left(-\varepsilon-\frac{\alpha}{1+\chi}\right)+f_p,\label{eq:uf2}\\
\chi&=&\frac{f_s}{\lambda_1}\left[1+\frac{\alpha}{\lambda_1(1+\chi)}\right]^{-1}\label{eq:uf3}.
\EE
These equations can be solved numerically to give $\{\chi,\lambda_1,M\}$ as functions of the model parameters, and we will compare them against numerical simulations in Section \ref{sec:pg}. Given the frozen nature of the dynamics in this phase, one finds $H=\sigma^2$, and a calculation similar to that of \cite{GallSher05,book2} gives the explicit form
\be
H=\sigma^2=\frac{1}{2}\frac{1+M}{(1+\chi)^2}.
\ee

The validity of Eqs. (\ref{eq:uf1},\ref{eq:uf2},\ref{eq:uf3}) breaks down when either $\chi\to\infty$ or $\lambda_1\to 0$. Assuming $\chi\to\infty$ and combining Eqs. (\ref{eq:uf1}) and (\ref{eq:uf3}) at $\varepsilon\neq 0$ implies a divergence of $f_s$, so that we can exclude ergodicity breaking at non-zero values of $\varepsilon$. We will therefore focus on $\varepsilon=0$ in the following.  Numerical solutions of the equations suggest a transition between an ergodic and non-ergodic phase as $n_s$ is raised, and that the product $\lambda_1\chi$ remains finite at this transition, while $\chi\to\infty, \lambda_1\to 0$.  Denoting the location of the onset of ergodicity-breaking by $n_s^c$, and using the ansatz of a finite limit $\lim_{n_s\to n_s^c}\lambda_1\chi$ one can multiply Eq. (\ref{eq:uf1}) by $\chi$ and Eq. (\ref{eq:uf3}) by $\lambda_1$ and take the limit $n_s\to n_s^c$. Writing $\alpha_c=1/(n_p+n_s^c)$ one finds
\begin{equation}
\alpha_c^2(2n_p-1)+2\alpha_c-(1-\alpha_c n_p)^2=0.
\label{eq:alphaquad}
\end{equation}
For $n_p=1$ this equation gives $\alpha_c=0.25$ which corresponds to $n_s^c = 3$, as we will confirm in simulations below. For a general value of $n_p$ one finds

\begin{equation}
\alpha_c=\frac{-(1+n_p)+2\sqrt{n_p}}{-n^2_p+2n_p-1},
\label{eq:alphaquadsolve}
\end{equation}
where a non-physical solution of Eq. (\ref{eq:alphaquad}) has been disregarded. Using $n_s=\alpha^{-1}-n_p$ this is easily converted into an explicit equation for $n_s^c$ as a function of $n_p$.
\subsection{Bounded ergodic phases}
In bounded phases one finds $\lambda(t)\equiv \lambda_0$ asymptotically in computer simulations, with $\lambda_0$ a finite constant (which depends on the choice of model parameters). Using this as an ansatz the analysis of (\ref{eq:1}-\ref{eq:3}) our calculation then proceeds by first transforming into Fourier space, and by an identification of the contributing modes. As in \cite{GallCoolSher03,GallSher05} one finds that the correlation function is of the form
\be
C(\tau)=c_0+(1-c_0)(-1)^\tau,
\ee
i.e. that only the Fourier modes $\omega=0$ and $\omega=\pi$ contribute (where the Fourier transform of $C(\tau)$ reads $\widetilde C(\omega)=\sum_\tau C(\tau)e^{-i\omega \tau}$, and similarly for $\widetilde G(\omega)$). Note that $\widetilde G(\omega=0)=\chi$ measures the response of the systems to persistent perturbations, while $\chi'\equiv G(\omega=\pi)=\sum_\tau (-1)^\tau G(\tau)$ indicates the reaction of the system to oscillatory perturbation fields. As in previous spherical MG models we find a phase in which only the mode at $\omega=0$ contributes to the correlation function, i.e. a fully frozen phase $C(\tau)\equiv 1$, and a second oscillatory phase in which both modes contribute. We will derive closed equations for the corresponding order parameters in the following. Similar to what was found in \cite{GallSher05,book2} the predictability $H$ and volatility $\sigma^2$ are given by
\BE
H&=&\frac{1}{4}\frac{1+2M+c_0}{(1+\chi)^2}, \label{eq:hh} \\
\sigma^2&=&H+\frac{1}{4}\frac{1-c_0}{(1+\chi')^2}.\label{eq:sigmasigma}
\EE
We note here that these expressions are exact results in the thermodynamic limit, with no approximations made. In particular Eq. (\ref{eq:sigmasigma}) provides an explicit and exact result for the market volatility of the model. No such exact expression is available in conventional MGs, where instead one is limited to approximations neglecting transient contributions to the dynamical order parameters (see \cite{book2} for details).
\subsubsection{Frozen phase with bounded dynamics (BF)}
In the frozen phase with bounded dynamics we have $\lambda(t)\equiv \lambda_0$ and $C(\tau)\equiv 1$. We can thus focus on the order parameters $\{\lambda_0, M,\chi\}$. The response $\chi'$ to alternating perturbations may be non-trivial, but is of no consequence for the observables in this phase. Performing a Fourier transform of (\ref{eq:1}-\ref{eq:3}) and subsequently focusing on the zero-frequency component one finds

\BE
M&=&f_p-f_s\left[\varepsilon+\frac{\alpha}{1+\chi}\right]\left[\rho\lambda_0+\frac{\alpha}{1+\chi}\right]^{-1},
\label{eq:BF_final_Me}\\
\rho\lambda_0&=&\left[-\left(\frac{M-f_p}{f_s}\right)\left(\varepsilon+\frac{\alpha}{1+\chi}\right)+ \frac{2\alpha\chi(1+M)}{f_s(1+\chi)^2}-\frac{\alpha}{1+\chi}\right],
\label{eq:BF_final_lambda0}\\
\chi&=&f_s\left(\rho\lambda_0 +\frac{\alpha}{1+\chi}\right)^{-1}.
\label{eq:BF_final_Chie}
\EE
These three equations are readily solved numerically to give $\{\lambda_0, M,\chi\}$ as a function of the model parameters $\{n_s,\rho,\varepsilon\}$. Near perfect agreement with numerical simulations will be reported in Section \ref{sec:pg}.  We here note that assuming a divergence of the susceptibility $\chi$ in Eq. (\ref{eq:BF_final_Chie}) directly leads to $\rho\lambda_0=0$ (provided we focus on non-trivial cases $f_s>0$). Inserting this in Eq. (\ref{eq:BF_final_Me}) then gives $\varepsilon=0$. Thus ergodicity breaking may occur in the BF phase at most if $\varepsilon=0$, and assuming that $\lambda_0$ remains positive at such a transition, a divergence of $\chi$ would also imply $\rho=0$. At $\varepsilon=\rho=0$ however, no BF phase is observed in simulations, so that we conclude that the model does not exhibit a BF$\to$ NE transition. 

\subsubsection{Oscillatory phase (O)}
An oscillatory phase at finite Lagrange multiplier $\lambda(t)\equiv\lambda_0$ is described by the following five equations, obtained from the Fourier transforms of (\ref{eq:1}-\ref{eq:3}) evaluated at $\omega=0$ and $\omega=\pi$ (one should note here that the one-time observable $M(t)$ becomes time-independent in the stationary state, so that Eq. (\ref{eq:1}) only needs to be considered at $\omega=0$):
\BE
M&=&f_p-f_s\left[\varepsilon+\frac{\alpha}{1+\chi}\right]\left[\rho\lambda_0+\frac{\alpha}{1+\chi}\right]^{-1},
 \\
\hspace{-5em}\rho\lambda_0\left(c_0-f_p\right)&=&-\left(\varepsilon+\frac{\alpha}{1+\chi}\right)(M-f_p)-\alpha\frac{c_0-f_p}{1+\chi}\nonumber\\
&&+\alpha\frac{1+c_0+2M}{(1+\chi)^2}\chi,\\
\hspace{-5em}\rho\lambda_0\chi&=&-\alpha\frac{\chi}{1+\chi}+f_s, \\
\hspace{-5em}(\rho-2)\lambda_0(1-c_0)&=&-\alpha\frac{1-c_0}{(1+\chi')}+\alpha\frac{(1-c_0)\chi'}{(1+\chi')^2},\label{eq:osc4}\\
\hspace{-5em}(\rho-2)\lambda_0\chi'&=&-\alpha\frac{\chi'}{1+\chi'}+f_s.\label{eq:osc5}
\EE
Since $c_0\neq 1$ in the oscillatory phase, the factor $1-c_0$ may be eliminated in Eq. (\ref{eq:osc4}), yielding closed equations (\ref{eq:osc4}, \ref{eq:osc5}) for $\chi'$ and $\lambda_0$, from which ones finds
\BE
\lambda_0&=&-\frac{n_s(1+\sqrt{\frac{1}{n_s}})^2}{(\rho-2)(n_p+n_s)},\\
\chi'&=&-\frac{1}{1+\sqrt{\frac{1}{n_s}}},
\EE
as the physically relevant solutions. These results can then be used to determine $\chi,c_0$ and $M$ from the remaining equations, and one as
\BE
\hspace{-4em}\chi&=&\frac{-(\rho\lambda_0+\alpha-f_s)+\sqrt{(\rho\lambda_0+\alpha-f_s)^2+4\rho\lambda_0f_s }}{2\rho\lambda_0},\label{eq:osc1}\\
\hspace{-4em}M&=&f_p-\frac{\epsilon+\frac{\alpha}{1+\chi}}{\rho\lambda_0 + \frac{\alpha}{(1+\chi)}}f_s,\label{eq:osc2} \\
\hspace{-4em}c_0&=&\frac{f_p\rho\lambda_0-(\epsilon+\frac{\alpha}{1+\chi})(M-f_p)+\alpha \frac{f_p}{1+\chi}+\alpha\frac{\chi(1+2M)}{(1+\chi)^2}}{\rho\lambda_0 + \frac{\alpha}{1+\chi}-\frac{\alpha\chi}{(1+\chi)^2}}.\label{eq:osc3}
\EE
The description of this phase breaks down whenever $c_0\to 1$ or $\chi\to \infty$ (or when no positive solution for $\lambda_0$ can be found). In all tested cases, the condition $c_0<1$ is the first one to be violated as $n_s$ is lowered, and a transition to a BF phase is found. The onset of this phase, as depicted in the phase diagrams presented in the following section, is identified numerically from the solutions of Eqs. (\ref{eq:osc1})-(\ref{eq:osc3}). 

This concludes the derivation of analytical expressions for the various order parameters in the different phases found in the model.
\section{Phase diagram and test against simulations: spherical GCMGs without memory loss}\label{sec:pg}
We now proceed to test the theoretical results obtained for the different regimes against numerical simulations and to work out the phase diagram of the spherical GCMG without score-memory loss. The setup in which agents are subject to exponential discounting will be considered in the next section. Unless indicated otherwise in figure captions, simulations are typically performed at system sizes of $PN_s=10^4$ (occasionally ranging up to $PN_s=2.56\cdot 10^5$), run for $100$ batch steps, and averaged over $10$ realisations of the random strategy assignments.

\begin{figure}[t!!!]
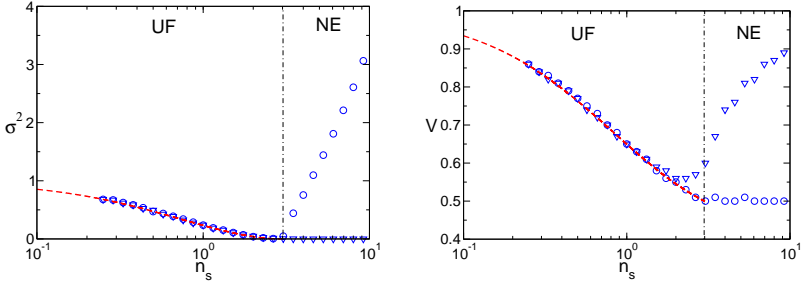

  \vspace*{10mm} \epsfxsize=50mm \epsffile{e0r0sigma.eps} ~~~~\epsfxsize=50mm \epsffile{e0r0V.eps}\vspace*{2mm}
  \caption{(Colour on-line) Spherical GCMG at $\varepsilon=0$ and without memory loss ($\rho=0$). Relative number of producers is $n_p=1$. The left panel shows the volatility $\sigma^2$ as a function of $n_s$, the right panel the traded volume $V$. A transition between an unbounded frozen phase (UF) at $n_s<n^c=3$ and a non-ergodic phase (NE) at $n_s>n^c=3$ is observed. Lines are the theoretical predictions in the UF phase, markers are from simulations, with circles corresponding to initial conditions with low bias ($u_0=0.01$) and squares to highly biased starts ($u_0=5$). Vertical lines mark the location of the phase transition as predicted from the theory}
\label{fig:e0r0}
\end{figure}

\begin{figure}[t!!!]
  \vspace*{10mm} 
\begin{center}\epsfxsize=50mm \epsffile{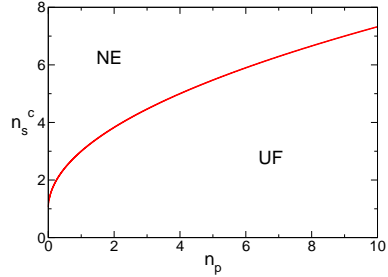} 
\end{center}
  \caption{Phase diagram of the model at $\varepsilon=0$ without memory loss ($\rho=0$). The line depicts the analytically obtained phase boundary separating the UF and NE phases as obtained from Eq. (\ref{eq:alphaquadsolve}).}
\label{fig:pge0r0}
\end{figure}

\begin{figure}[t!!!]
\begin{center}
  \vspace*{10mm} \epsfxsize=50mm \epsffile{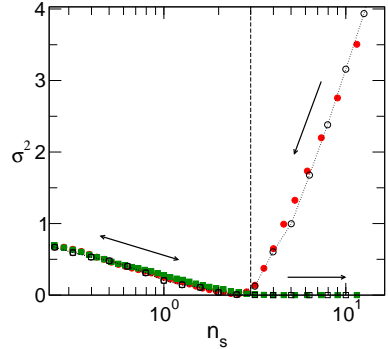} 
\end{center}
  \caption{(Colour on-line) Hysteresis in the spherical GCMG at $\rho=\epsilon=0$ (filled symbols). Simulations are started at high $n_s$ and low initial bias (high volatility branch), and $n_s$ is then reduced adiabatically at fixed $n_p=1$ (filled circles) across the phase transition of the equilibrium model, and then increased again (filled squares). The system then enters the low volatility branch at $n_s>n_s^c$. Open symbols show equilibrium results for comparison. Adiabatic variation of $n_s$ is performed at fixed $N=200$, varying $P=N_p$ between values of $16$ and $160$, measurements are taken when the volatility varies by no more than $0.1$ per cent from one batch step to the next. Averages over $10$ samples are taken. }
\label{fig:hyster}
\end{figure}

\subsection{The case $\varepsilon=0$, $n_p>0$}
Results for the case $\rho=\varepsilon=0$ at a non-zero fraction of producers are depicted in Figs. \ref{fig:e0r0}, \ref{fig:pge0r0} and \ref{fig:hyster}. As in the conventional GCMG one finds a transition between an ergodic phase at low relative numbers of speculators $n_s<n_s^c$, and a non-ergodic phase in which the stationary state depends on initial conditions at $n_s>n_s^c$. More precisely the ergodic regime is of the bounded frozen type (UF) as described in the previous section. Numerical measurements of order parameters such as the volatility and the traded volume agree well with theoretical predictions in this ergodic regime, as seen in Fig. \ref{fig:e0r0}. The location of the phase transition is obtained from Eq. (\ref{eq:alphaquad}), and we note that $n_s^c=1/\alpha_c-n_p$. In particular one finds $n_s^c=3$ for $n_p=1$. The behaviour of $n_s^c$ as a function of $n_p$ is depicted in the phase diagram shown in Fig. \ref{fig:pge0r0}, and the numerical value of $n_s^c$ is found to increase as $n_p$ increases.

In Fig. \ref{fig:hyster} we finally illustrate a further interesting dynamical effect in the spherical GCMG at $\rho=\varepsilon=0$ and show that the system exhibits hysteresis if the fraction of speculators is varied `adiabatically' across the ergodicity-breaking phase transition. A similar effect has been reported for the conventional MG in \cite{moro}. In particular we start our simulations at large values of $n_s$ from unbiased initial conditions, i.e. in the high-volatility branch of the non-ergodic phase. $n_s$ is then gradually reduced and swept across the transition point $n_s^c$. No re-initialisation of the strategy scores is performed at any point. As seen in Fig. \ref{fig:hyster} the volatility measured during this process follows that of corresponding equilibrium simulations at the same model parameters. Far below the transition (i.e. at an $n_s<n_s^c$) we then revert the dynamics, and start to increase $n_s$ adiabatically, again crossing the phase transition point. At $n_s>n_s^c$ the system then enters the low volatility branch of the non-ergodic phase, i.e. the initial starting point of on the high-volatility branch has been washed out of the dynamics, and no memory of these initial conditions appear to be seen in the macroscopic dynamics.

\subsection{The case $\varepsilon=n_p=0$}

GCMGs without memory-loss and operated at $n_p=\varepsilon=0$ have been reported to have somewhat different dynamical characteristics than GCMGs at non-zero numbers of producers ($n_p>0)$. In particular, as described in \cite{book2}, agents are found to stop trading (i.e. the traded volume is $V=0$), even though their strategy scores do not diverge linearly in time. In this sense speculators are not frozen in the common language of MG theory, and susceptible to persistent perturbations in the strategy scores \cite{book2,tonprivate}. An ergodicity-breaking phase transition can hence still occur, even though the macroscopic dynamics in the ergodic regime may appear trivial given that all players retire from trading.
\begin{figure}[t!!!]
\begin{center}
  \vspace*{10mm} \epsfxsize=50mm \epsffile{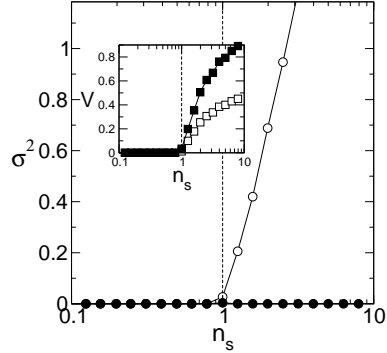} 
\end{center}
  \caption{Behaviour of the model without memory-loss at $\varepsilon=n_p=0$. The figure shows the volatility (main panel) and the traded volume (inset) as measured in simulations ($PN_s=10^4$, run for $2000$ batch steps, averages over $10$ samples). Open symbols indicate initial conditions with small bias ($u_0=0.01$), filled markers correspond to biased starts ($u_0=5$). The system is in the BF phase below the phase transition at $n_s^c=1$ (vertical line), and non-ergodic above $n_s^c$ (see text).}
\label{fig:np0eps0}
\end{figure}
We have probed the behaviour of the spherical GCMG at $n_p=\varepsilon=0$, always restricting the discussion to the case without memory-loss ($\rho=0$). Results from simulations are presented in Fig. \ref{fig:np0eps0}. One observes ergodicity breaking at large $n_s$, and vanishing volatility and trading volume in the ergodic phase, i.e. as in the conventional GCMG agents stop trading at $n_p=\varepsilon=0$ in the ergodic phase. Further inspection of our simulations reveal that strategy scores remain negative, but finite in the ergodic phase, leading us to the conclusion that the behaviour below the transition is of the bounded frozen (BF) type discussed above. Setting $\rho=n_p=\varepsilon=0$ in Eq. (\ref{eq:BF_final_Me}) directly leads to $M=-1$ (note that $n_p=0$ implies $f_s=1,f_p=0$), in agreement with the results from simulations. Inserting $c_0=1$ and $M=-1$ in Eq. (\ref{eq:hh},\ref{eq:sigmasigma}) then gives $\sigma^2=0$ as found in simulations. Setting $\rho=0$ in Eq. (\ref{eq:BF_final_Chie}), and assuming that $\lambda_0$ is finite, furthermore gives $\chi=(\alpha-1)^{-1}$, i.e. the susceptibility diverges at $\alpha_c=1$, or equivalently we have $n_s^c=\alpha_c^{-1}-n_p=1$, again fully consistent with the results from simulations shown in Fig. \ref{fig:np0eps0} and with the phase diagram shown in Fig. \ref{fig:pge0r0}.

\subsection{The case $\varepsilon\neq 0$}
\begin{figure}[t!!!]
  \vspace*{10mm} \epsfxsize=50mm \epsffile{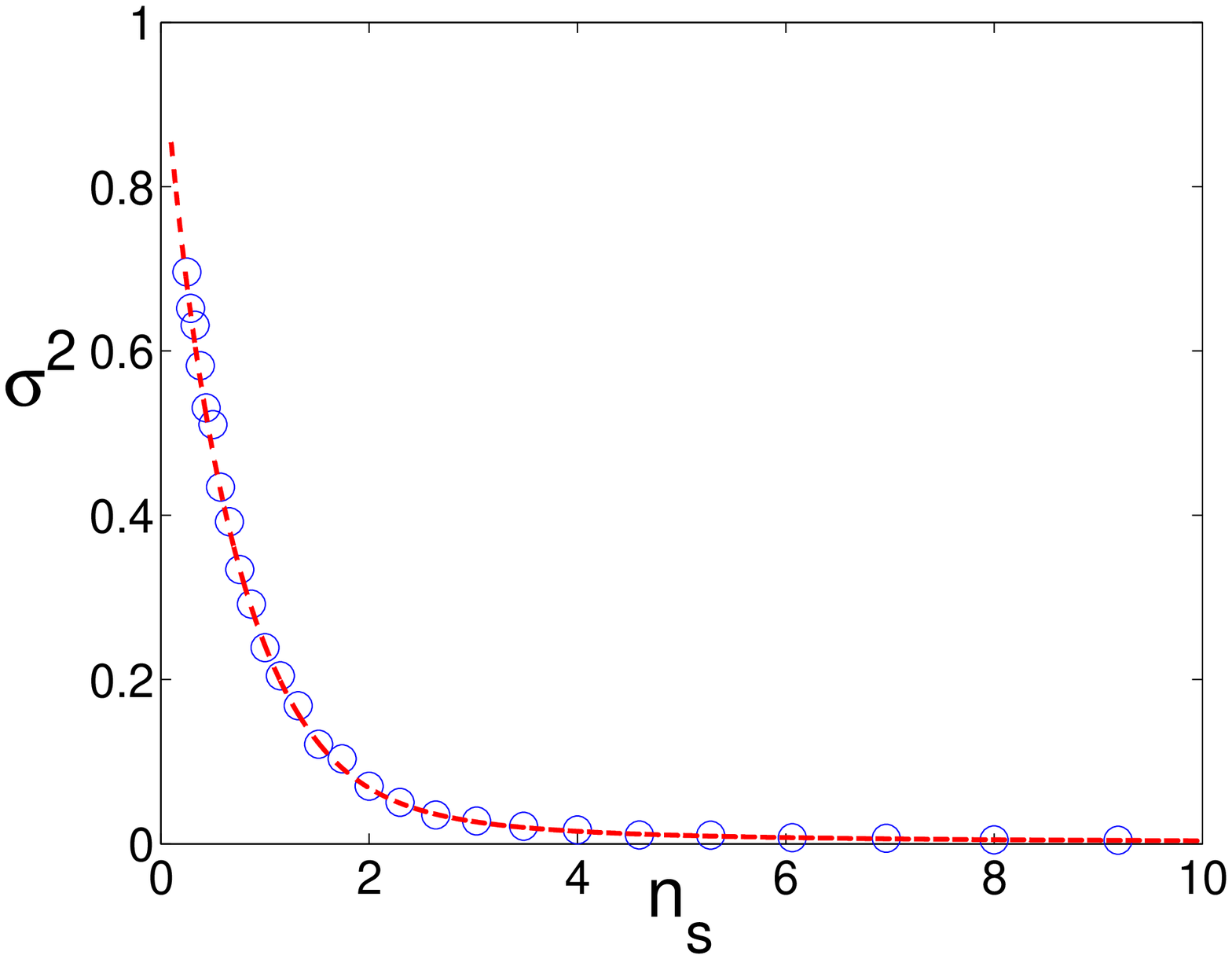}  ~~~~~~ \epsfxsize=47mm \epsffile{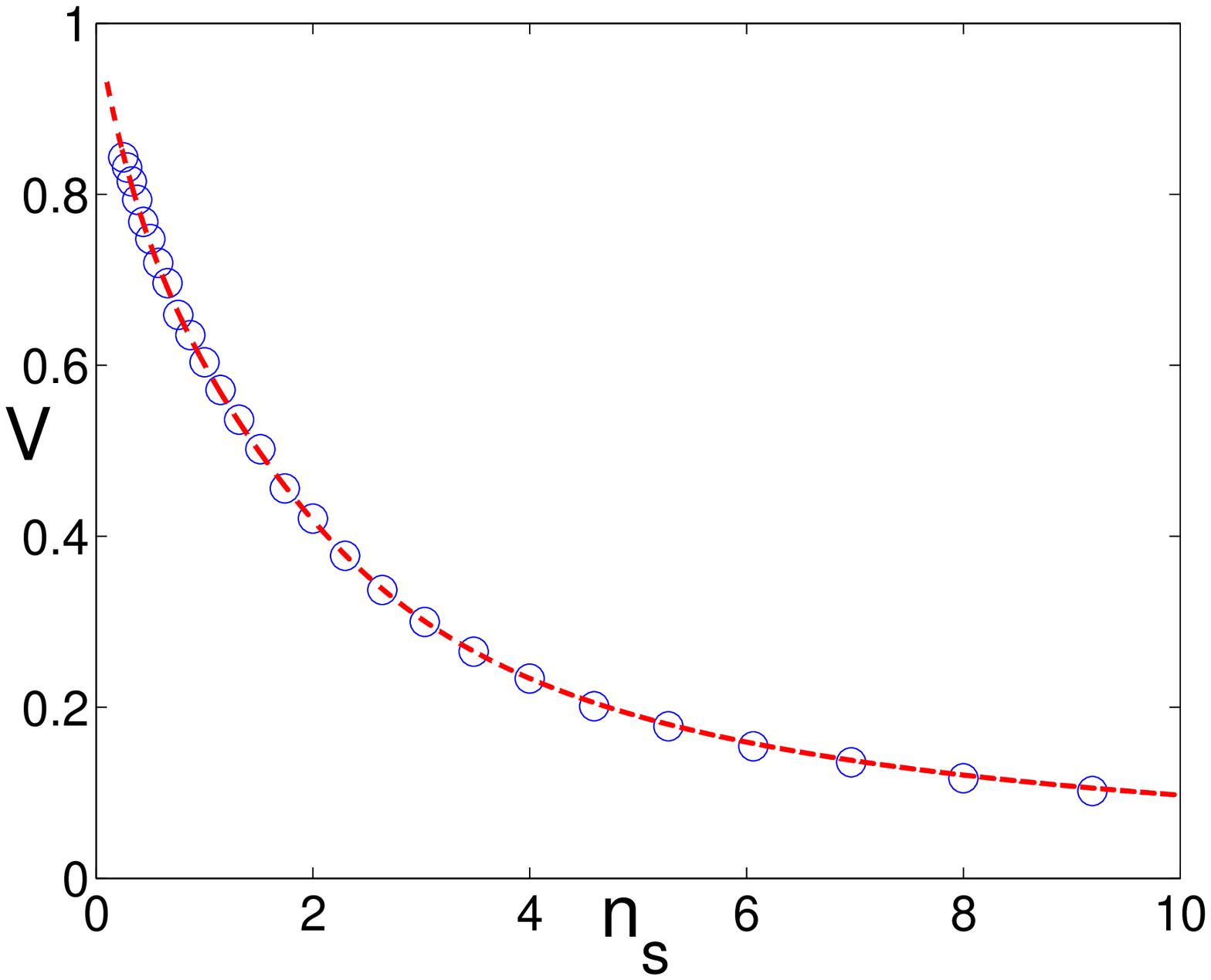} 
  \caption{(Colour on-line). Model without memory loss ($\rho=0$), but at non-zero disincentive to trade ($\varepsilon=0.2$). No phase transition is observed, and the system is in the UF phase for all $n_s>0$. The figure shows the volatility (left panel) and the traded volume (right panel), with markers from simulations and lines from the analytical predictions for the UF phase.}
\label{fig:e2r0}
\end{figure}

The case $\rho=0$ and $\varepsilon\neq 0$ is relatively unspectacular, no phase transition is found similar to what is seen in the conventional GCMG \cite{book2}. Simulation results for the case $\rho=0$, $\varepsilon=0.2$ are shown in Fig. \ref{fig:e2r0}. The system exhibits an unbounded frozen phase for all values of $n_s$. Results of simulation show good agreement with the theoretical prediction, and are here reported mostly for completeness. 

\section{Spherical GCMGs with memory loss}\label{sec:finitemem}
\subsection{Uniform memory loss rates}
\begin{figure}[t]
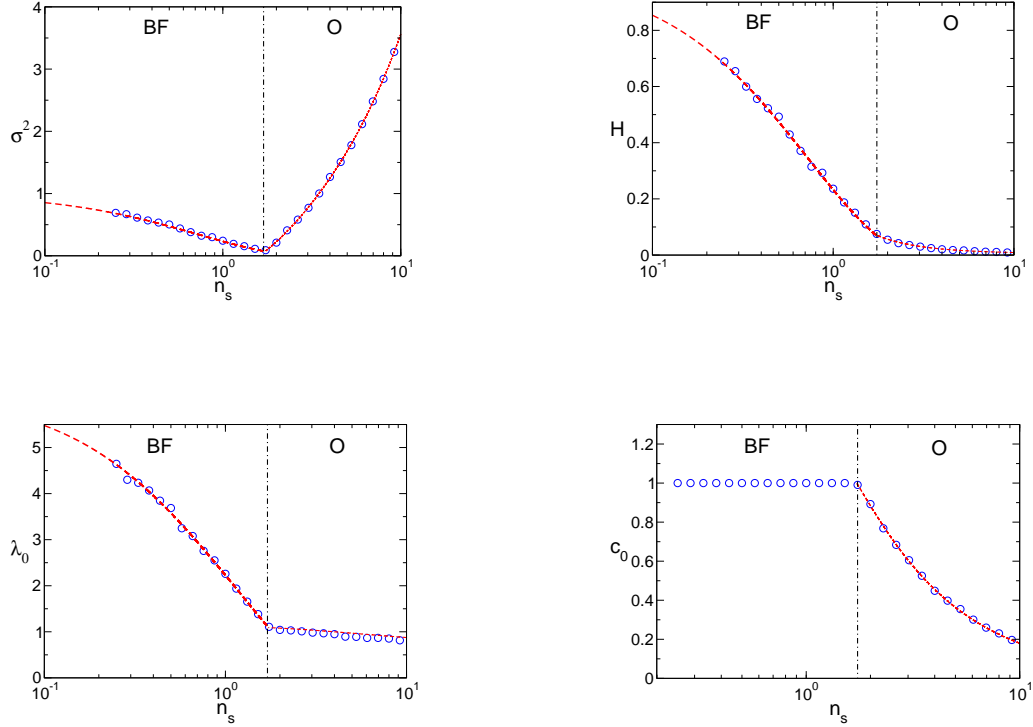

\vspace*{-0mm} \hspace*{5mm} \setlength{\unitlength}{1.0mm}
\begin{tabular}{cc}
\begin{picture}(100,55)
\put(-8,5){\epsfysize=40\unitlength\epsfbox{e0r2_sigma.eps}}
\end{picture} ~~~~~~~~ & 
\begin{picture}(100,55)
\put(-43,5){\epsfysize=40\unitlength\epsfbox{e0r2_H.eps}}
\end{picture}
\\
\begin{picture}(100,55)
\put(-8,5){\epsfysize=40\unitlength\epsfbox{e0r2_lambda.eps}}
\end{picture} ~~~~~~~~ & 
\begin{picture}(100,55)
\put(-43,5){\epsfysize=40\unitlength\epsfbox{e0r2_C.eps}}
\end{picture}
\end{tabular}
\vspace*{5mm}
  \caption{(Colour on-line) Spherical GCMG with memory loss ($\varepsilon=0, \rho=0.2, n_p=1$). The four panels show the volatility $\sigma^2$ (top left), the predictability $H$ (top right) as well as the Lagrange multiplier $\lambda_0$ (bottom left) and the persistent part of the correlation function $c_0$ (bottom right) as functions of $n_s$. A phase transition between a bounded frozen phase (BF) at $n_s<n_s^*$ and an oscillatory phase (O) at $n_s>n_s^*$ is observed as indicated by the vertical dashed lines. Good quantitative agreement between theory (lines) and numerical simulations (markers) is found, even though finite-size and equilibration effects yield slight deviations of measurements for $\lambda_0$ from the theoretical predictions. }
\label{fig:e0r2}
\end{figure}

\begin{figure}[t!!!]
  \vspace*{10mm} 
\begin{center}
\epsfxsize=50mm\epsffile{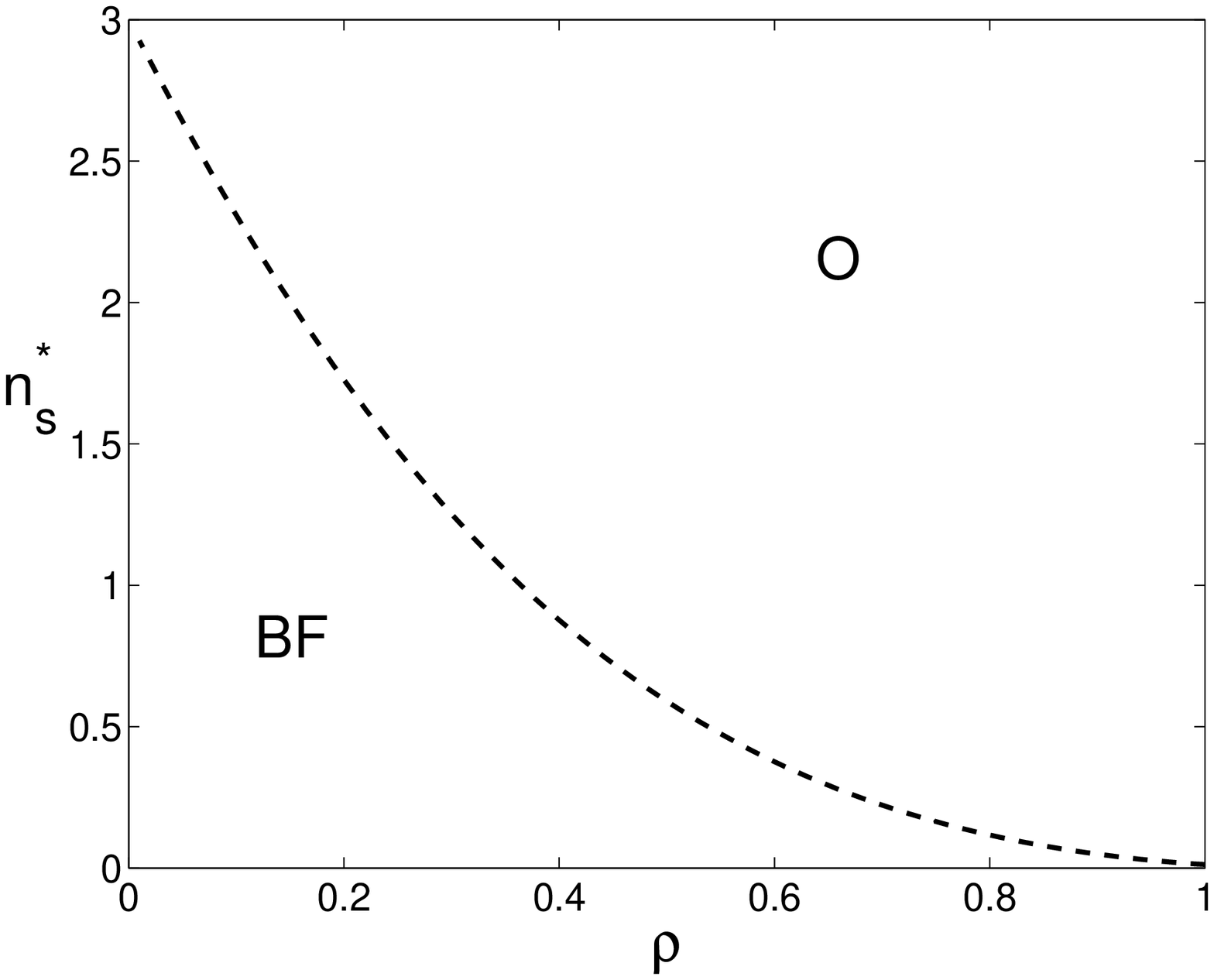} 
\end{center}
  \caption{Phase diagram for the model at uniform loss rate and at vanishing disincentive to trade ($\varepsilon=0$). The line separating the oscillatory and bounded frozen phase is obtained by solving the order parameter equations in the oscillatory phase, and by identifying the point at which $c_0\uparrow 1$ as $\rho$ is lowered. Labels O and BF are accurate only so long as $\rho>0$. On the line $\rho=0$ in parameter space the system is in an UF phase at $n_s<n_s^\star=3$ and in the non-ergodic regime at $n_s>3$ ($n_p=1$).}
\label{fig:uniform}
\end{figure}

The behaviour of the spherical GCMG at uniform memory-loss rate ($\rho>0$) and $\varepsilon=0$ is illustrated in Fig. \ref{fig:e0r2}. We here find no signs of ergodicity breaking and instead observe a bounded frozen phase at $n_s<n_s^*$ and an oscillatory one at $n_s>n_s^*$. As seen in the figure, numerical simulations confirm the theoretical predictions convincingly. The location of the phase transition between the oscillatory and bounded frozen phase is obtained from the numerical solution of Eqs. (\ref{eq:osc1}-\ref{eq:osc3}) as the point where $c_0\uparrow 1$. Plotting the so-obtained $n_s^*$ as a function of the memory-loss rate $\rho$ yields the phase diagram depicted in Fig. \ref{fig:uniform}. In particular no BF phase is found at complete memory loss ($\rho=1$), and that the system is fully oscillatory in this case. The limit $\rho\to 0$ on the other hand gives $n_s^*\to 3$ at $n_p=1$, i.e. the location $n_s^*$ of the O/BF transition approaches that of the UF/NE transition found for the model at $\varepsilon=\rho=0$, as discussed in the previous section. It is here appropriate to mention that the model at $\rho\neq 0, \varepsilon\neq 0$ is found to be in a BF phase in all tested cases, with perfect agreement between theoretical predictions and numerical simulations (not reported here). 

\subsection{Heterogeneous memory loss rates and existence of the efficient phase}

We will now allow agents to operate at heterogeneous,
i.e. player-dependent memory loss parameters. We will limit this
heterogeneity to the case where a fraction of speculators, $q$, have
$\rho=0$ (i.e. they operate at perfect memory capacities) and where
the remaining speculators have a uniform non-zero memory-loss rate $\rho$
(i.e. they are subject to exponential discounting). We also restrict the discussion to the case $\varepsilon=0$. Simulations (at $q<1$ reveal
that this case gives rise to an ergodic unbounded phase for
$n_s<n_s^c$, and a non-ergodic phase for $n_s>n_s^c$, where $n_s^c$
will generally depend on $n_p,q$ and $\rho$. As throughout most of
this paper we limit the discussion to the case $n_p=1$.  For values of
$n_s$ that present unbounded dynamics, one finds in numerical
experiments that the scores of agents with perfect memory increase
linearly with time whereas those of the `forgetful' agents approach finite values. Since in an unbounded frozen phase $\lambda(t)=\lambda_1
t$, only the perfect-memory agents will have a
$\phi_i(t)=\frac{u_i(t)}{\lambda(t)}$ that remains non-zero asymptotically. Hence, in
calculating the dynamics for the ergodic phase we will apply the
ansatz that there is only a contribution from the agents with $\rho=0$ to the relevant order parameters, i.e. in particular we start from the assumption that $C_s(\tau)=qC^{\rho=0}_s(\tau)$, $M_s =
qM^{\rho=0}_s$ and $G_s(\tau) =
qG^{\rho=0}_s(\tau)$, where e.g. $C_s^{\rho=0}$ is the correlation function obtained from considering all speculators who operate at $\rho>0$, and similarly for $M_s^{\rho=0}$ and $G_s^{\rho=0}$. Proceeding with the same method used in
the homogeneous case one then finds the following relations for the relevant persistent order parameters:
\BE
\chi&=&\frac{qf_s}{\lambda_1}\left[1+\frac{\alpha}{\lambda_1}\frac{1}{1+\chi}\right]^{-1}, \label{eq:hetero_chi}\\
M&=&\frac{qf_s}{\lambda_1}\left[1+\frac{\alpha}{\lambda_1}\frac{1}{1+\chi}\right]^{-1}\left(-\frac{\alpha}{1+\chi}\right)+f_p,\label{eq:hetero_M}\\
\lambda_1&=& \sqrt{q}\sqrt{\frac{\left(\frac{\alpha}{1+\chi}\right)^2 + \alpha\frac{2(1+M)}{(1+\chi)^2}}{\left( 1+\frac{\alpha}{\lambda_1}\frac{1}{1+\chi} \right)^2}}.
\label{eq:hetero_lambda}
\EE

\begin{figure}[t!!!]
  \vspace*{10mm} 
\begin{center}
\epsfxsize=50mm \epsffile{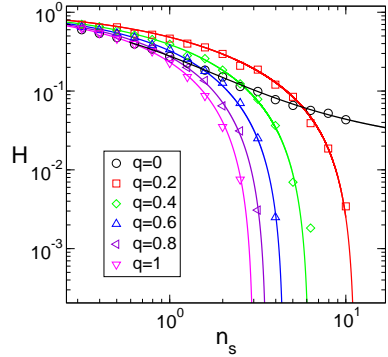} 
\end{center}
  \caption{(Colour on-line). Predictability versus $n_s$ at different fractions $q$ of agents which experience no memory loss (the remaining $(1-q)N_s$ speculators operate at memory-loss rate $\rho=0.5$). The figure depicts analytical results (lines) and measurements from simulations (markers) of the spherical GCMG with discounting. A transition to efficiency is observed at any $q>0$, but not if there are no agents operating at zero memory-loss rate. There instead a BF$\to$O transition is found. Simulations are performed at $PN_s=10^5$, run for $3000$ batch steps to minimise finite-size effects and to ensure equilibration }
\label{fig:het}
\end{figure}

\begin{figure}[t!!!]
  \vspace*{10mm}
  \begin{center}\epsfxsize=50mm \epsffile{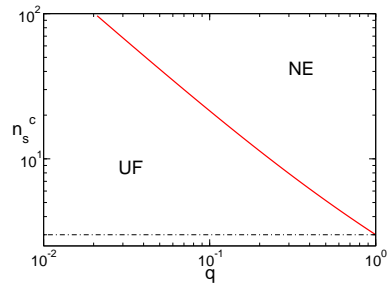}
\end{center} 
  \caption{(Colour on-line). Phase diagram for the model with heterogeneous memory-loss rates. A fraction of $q$ agents has zero memory loss rate, the precise value is irrelevant (see text).The upper line, separating the NE and UF phases, is the solution given by Eq. (\ref{eq:qphase}). The horizontal line is $n_s^c=3$, the result obtained in the limit $q\to 1$ (all agents operate at zero memory-loss). }
\label{fig:hetpg}
\end{figure}

These equations are again solved numerically. We compare the so-obtained analytical predictions against numerical simulations in Fig. \ref{fig:het}. Results show good agreement, even though the region at large $n_s$ (i.e. deep into the non-ergodic phase) is hard to probe in simulations due to finite-size effects.  The key result is the observation that an information efficient phase ($H=0$) exists only when the system has a finite fraction of perfect memory agents, i.e. when $q>0$\footnote{Interestingly, we observe that the predictability at $q=0$ agrees perfectly with that for the case $q=1$ at approximately $n_s<0.59$. Here, the system at $q=0$ is in the BF phase, whereas the system at $q=1$ is in the UF phase. Straightforward algebraic manipulations in fact show that Eqs. (\ref{eq:BF_final_Me},\ref{eq:BF_final_lambda0},\ref{eq:BF_final_Chie}) are equivalent to Eqs. (\ref{eq:uf1},\ref{eq:uf2},\ref{eq:uf3}) if $\rho\lambda_0$ is identified with $\lambda_1$. They hence yield identical results for $M$ and $\chi$, and the resulting expressions for the predictability in the two frozen phases are identical.}. This is confirmed analytically by identifying the onset of ergodicity breaking $\chi\to\infty$ from Eqs. (\ref{eq:hetero_chi}-\ref{eq:hetero_lambda}).  For simplicity, we now restrict the further discussion to $n_p=1$. Following the same steps as in the derivation of (\ref{eq:alphaquad}) one finds
\begin{equation}
\alpha_c^2(q-1)-2\alpha_c(q+1)+q= 0,
\label{eq:qquad}
\end{equation}
for the case of heterogeneous memory-loss rates. For $q=1$ this reduces to Eq. (\ref{eq:alphaquad}). Solving Eq. (\ref{eq:qquad}) then gives

\begin{equation}
\alpha_c=\frac{(1+q)\pm\sqrt{1+3q}}{(q-1)},
\label{eq:qphase}
\end{equation}

where the negative root corresponds to the physically relevant solution. Fig. \ref{fig:hetpg} then shows the resulting phase diagram in the $(q,n_s)$ plane, and in particular one finds that $n_s^c\to\infty$ as $q\to 0$, i.e. no non-ergodicity breaking is found for $q=0$. Equivalently we can conclude from these findings that the efficient non-ergodic phase at zero predictability can only be achieved if a non-zero fraction of agents uses an `optimal' learning rule to take their trading decisions, i.e. one with $\varepsilon=0$ and without memory-loss. Similar results have been found for MGs with impact correction and heterogeneous resource levels in \cite{GallDeMa}. A discussion of the case of conventional GCMGs can be found in \cite{GallDeMa} as well.
 
\section{Concluding remarks}
In summary, we have presented a detailed analysis of the stationary
states of a spherical version of the grand-canonical minority
game. The model is designed to be fully solvable in the ergodic
regime, even at non-zero memory loss rates where conventional MGs have
resisted any analytical progress so far. We have shown that the spherical model
nevertheless exhibits complex phase behaviour. In particular,
similar to the conventional GCMG, a phase transition between an
ergodic regime at positive predictability and a non-ergodic phase and
vanishing predictability is found when agents are not subject to
score-memory loss and when the disincentive to trade is set to
zero. This transition and with it the efficient phase is absent at any
$\varepsilon\neq 0$ or if agents experience memory loss. We have also studied mixed populations, in
which a fraction $q$ of speculators operates a optimal learning rule
(i.e. this group is not subject to memory-loss) and where the
remaining $(1-q)N_s$ speculators are subject to exponential
discounting of score-valuations (i.e. they have $\rho>0$). The
efficient phase at $H=0$ is found to exist if and only if $q>0$. In
particular the critical value $n_s^c(q)$ separating the ergodic
inefficient regime ($n_s<n_s^c(q)$) from the efficient non-ergodic
phase at $n_s>n_s^c(q)$ can be shown analytically to tend to infinity
as $q\to 0$, i.e. $\lim_{q\downarrow 0}n_s^c(q)=\infty$. All analytical findings have been verified successfully in numerical simulations.
The spherical model necessarily represents a substantial reduction in complexity as compared to non-spherical GCMGs and stylised facts are presumably much less likely to be found in the spherical model. Due to its analytical tractability, especially in cases with memory-loss, the spherical model as discussed here might however serve as a starting point for further studies of MGs with score-discounting, for example within a perturbative expansion taking into account non-linear corrections to the effectively harmonic update rules of the spherical model.
\section*{Acknowledgements}
This work is supported by an RCUK Fellowship (RCUK reference EP/E500048/1). TG has the pleasure of thanking ACC Coolen and D Sherrington for earlier collaboration on spherical minority games. 


\end{document}